\documentclass[traditabstract]{aa} 
\usepackage{graphicx}
\usepackage{natbib}
\bibpunct{(}{)}{;}{a}{}{,}
\usepackage{epstopdf}
\usepackage{color}
\usepackage{txfonts}
\usepackage{hyperref}
\newcommand{\teff}{$T_{\rm eff}$}
\newcommand{\logg}{$\log g$}
\newcommand{\mlsep}{$\langle \Delta \nu \rangle$}
\newcommand{\numax}{$\nu_{\rm max}$}
\newcommand{\pmm}{$\pm$}
\newcommand{\msol}{M$_{\odot}$}
\newcommand{\rsol}{R$_{\odot}$}
\newcommand{\lsol}{L$_{\odot}$}

\newcommand{\zx}{$Z_{\rm i}/X_{\rm i}$}
\newcommand{\mhz}{$\mu$Hz}
\newcommand{\gmb}{Gmb\,1830}
\begin{document}
   \title{Fundamental properties of the Population II fiducial stars \object{HD\,122563} and \object{Gmb\,1830} from CHARA interferometric observations}

   \author{O.~L.~Creevey\inst{1},
          F. Th\'evenin\inst{1},
          T.~S.~Boyajian\inst{2,3},
          P.~Kervella\inst{4},
          A. Chiavassa\inst{5},
          L. Bigot\inst{1},
          A. M\'erand\inst{6},
          U. Heiter\inst{7},
          P. Morel\inst{1},
          B. Pichon\inst{1},
          H. A. Mc Alister\inst{2},
          T. A. ten Brummelaar\inst{2},
          R. Collet\inst{8,9},
          G.~T.~van Belle\inst{10}
          V. Coud\'e du Foresto\inst{4},
          C. Farrington\inst{2},
          P. J. Goldfinger\inst{2},
          J. Sturmann\inst{2},
          L. Sturmann\inst{2},
          N. Turner\inst{2}
          }

   \institute{Laboratoire Lagrange, UMR 7293, CNRS, Observatoire de la C\^ote d'Azur, Universit\'e de Nice Sophia-Antipolis, Nice, France
              \email{ocreevey@oca.eu}
\and
Center for High Angular Resolution Astronomy, Georgia State University, PO Box 3965, Atlanta, Georgia 30302-3965, USA
\and 
Hubble Fellow
\and
LESIA-Observatoire de Paris, CNRS UMR 8109, UPMC, Univerit\'e Paris Diderot, 5 place Jules Janssen, F-92195 Meudon, France
         \and
Institut d'Astronomie et d'Astrophysique, Universit\'e Libre de Bruxelles,
CP. 226, Boulevard du Triomphe, 1050 Bruxelles,
Belgium
\and
European Southern Observatory, Alonso de C\'ordova 3107, Casilla 19001, Santiago 19, Chile
\and
Department of Physics and Astronomy, Uppsala University, Box 516, SE-75120 Uppsala, Sweden
\and
Centre for Star and Planet Formation, Natural History Museum of Denmark, University of Copenhagen, {\O}ster Voldgade 5-7, 1350 Copenhagen, Denmark 
\and
Astronomical Observatory/Niels Bohr Institute, Juliane Maries Vej 30, 2100 Copenhagen, Denmark    
\and
Lowell Observatory, 1400 West Mars Hill Road, Flagstaff, Arizona, 86001, USA
}

   \date{Received January 2012; accepted March 16, 1997}

 
  \abstract
   {
We have determined the angular diameters of two metal-poor stars, \object{HD\,122563} and
\object{\gmb},
using CHARA and Palomar Testbed Interferometer observations.  
For the giant star HD\,122563, 
we derive an angular diameter $\theta_{\rm 3D} = 0.940 \pm 0.011$ 
milliarcseconds (mas) 
using limb-darkening
from 3D convection simulations 
and 
for the dwarf star Gmb\,1830 (HD\,103095) we obtain a 1D limb-darkened 
angular diameter $\theta_{\rm 1D} = 0.679 \pm 0.007$ mas.
Coupling the angular diameters with photometry yields 
effective temperatures with
precisions better than 55 K 
(\teff\ = 4598 \pmm\ 41 K and 4818 \pmm\ 54 K --- 
for the giant and the dwarf star, respectively).
Including their distances results in very 
well-determined luminosities and radii 
($L = 230 \pm 6$ \lsol, $R = 23.9 \pm 1.9$ \rsol\ 
and $L = 0.213 \pm 0.002$ \lsol, $R = 0.664 \pm 0.015$ \rsol, respectively).
We used the CESAM2k stellar structure and evolution code in order to 
produce models that fit the observational data.  
We found values of the mixing-length parameter $\alpha$ (which describes
1D convection) that depend on the mass of the star.
The masses were determined from the models 
with precisions of $<$3\% and with the well-measured
radii excellent constraints on the surface gravity 
are obtained
($\log g = 1.60 \pm 0.04, 4.59\pm 0.02$ dex, respectively).
The very small errors on both $\log g$ and \teff\ 
provide stringent constraints for
spectroscopic analyses given the sensitivity of abundances to both of these 
values.
The precise determination of \teff\ for the two stars brings into
question the photometric scales for metal-poor stars.}

   {}
   {}{}{}


\authorrunning{Creevey et al.}
\titlerunning{Stellar properties of metal-poor stars from interferometry}

   \maketitle
%

\section{Introduction}
Metal-poor stars are some of the oldest stars in the Galaxy and 
thus reflect the chemical composition of Galactic matter
at the early stages of Galactic evolution.
The determination of accurate {\it observed} fundamental properties,
and in particular their location in the Hertzsprung-Russell (HR) diagram, is a key requirement 
if we aim to constrain the {\it unobservable} properties such as mass, age, 
and initial helium content by using stellar models.
Among the most controversial {\it observed} parameter is the 
effective temperature ($T_{\rm eff}$) 
which can vary by more than 200 K for metal-poor stars from one method
to another (see the PASTEL catalogue, \citealt{soubiran10}). 
In particular, local thermodynamic equilibrium (LTE) is usually assumed
and non-LTE (NLTE) 
effects must be included in spectroscopic analyses
especially for metal-poor stars where these effects are
enhanced \citep{thev99,andrie10,merle11}
and this leads to even more discrepancy between literature values.
One solution is to measure the angular diameter and convert this to \teff\ 
to provide a {\it direct} determination.

The large majority of metal-poor stars belong to the halo or the old disk of 
the Galaxy which means that their apparent magnitude and or angular diameters 
are extremely small and difficult to measure.
However, some instruments, in particular those on the CHARA array 
\citep{tenbrummelaar05}
are very capable of working at short wavelengths 
on long baselines to obtain the required angular resolution.
Among the most exciting possible targets with CHARA working in the $K$ band are 
HD 122563 (=~\object{HR 5270}, \object{HIP 68594}, m$_V$ = 6.19 mag)
and \gmb\
(=~\object{HD\,103095}, \object{LHS 44}, \object{HIP 57939}, m$_V$ = 6.45 mag)
whose mean metallicities $[Z/X]_s$\footnote{$[Z/X] = \log Z/X_{\rm star} - 
\log Z/X_{\odot}$ and Z/X$_{\odot} = 0.0245$, see Sect.~\ref{sec:models}} 
are $\sim$--2.3 dex 
and --1.3 dex, respectively (see discussion in Sect.~\ref{sec:fparamsobs}), where 
$Z$ and $X$ denote the metallicity and hydrogen (absolute) 
mass fraction in the star
and the subscript refers to the 
observed surface value.

HD 122563, a standard example of a very metal-poor field 
giant \citep{wallerstein63, wolffram72},  
has been extensively studied
and presents similarities with metal-poor giants found in globular clusters.
\gmb\ is a metal-poor halo dwarf star 
recognized as exhibiting depleted Li \citep{deli94,king97}
when compared to 
the mean value of halo dwarf stars \citep{spite93, ryan05}.
It is also the nearest halo dwarf and has an excellent parallax measurement. 
Combining interferometric measurements of these stars with other 
already measured old moderately metal-poor stars, 
such as $\mu$ Cas ([$Z/X$]$_s$~=~-0.5 dex, \citealt{boyajian08}), 
offers an excellent 
opportunity to constrain the \teff\ scale of metal-poor 
stars over a wide range of 
metallicities with possible implications for \teff\ calibrations of
globular cluster stars. 
In Table \ref{tab:teffs} we summarize some of the most recent determinations
of the atmospheric properties of both targets.
Note that HD~122563 and \gmb\ have 
also been defined as benchmark stars for the Gaia mission under
the SAM\footnote{\url{www.anst.uu.se/ulhei450/GaiaSAM/}} working group.

Not only are temperature scales for metal-poor stars controversial, but
stellar structure and evolution 
models often predict higher \teff\ than those observed for these stars 
(see e.g. Fig.~2 of \citealt{leb00}).  The difficulty encountered when 
trying to match evolutionary tracks to the observational data not only 
severely inhibits the determination of any fundamental properties 
but any chance of improving or testing the physics in the models is 
also limited.

Considering the difficulties mentioned above, 
in this paper we aim to determine accurate fundamental properties of 
HD\,122563 and \gmb\ based on interferometric observations
(Sect.~\ref{sec:observations}).
In Sect.~\ref{sec:diameters} we present our analysis of the observations
to determine the angular diameters of both stars.
We then determine the {\it observed} values of 
\teff, luminosity $L$, and radius $R$, and 
subsequently use  
stellar models to constrain the {\it unobservable} properties
of mass $M$, initial metal and helium content $Z_{\rm_i},Y_{\rm i}$, 
mixing-length parameter $\alpha$ 
and age (Sect.~\ref{sec:fparams}).
We also 
predict their global asteroseismic properties in order to determine
if such observations could further constrain the models.

\begin{table}
\begin{center}
\caption{Most recent photometric and spectroscopic determinations of 
atmospheric parameters for the target stars {\label{tab:teffs}}}
\begin{tabular}{lllllllll}
\hline\hline
\multicolumn{4}{l}{HD\,122563} & \multicolumn{4}{l}{\gmb}\\
\teff&\logg&[Fe/H]&p/s$^a$&
\teff&\logg&[Fe/H]&p/s\\
(K) & (dex) & (dex) & &
(K) & (dex) & (dex) & \\
\hline
$4795^b$& \dotfill & \dotfill & p & 5129$^b$ & \dotfill&\dotfill&p\\
4598$^c$& \dotfill & \dotfill  & p& 5011$^c$& \dotfill&\dotfill&p\\
4572$^d$& \dotfill & \dotfill & p&5054$^e$ &\dotfill&\dotfill&p\\
4600$^f$& 1.50 & -2.53 & s&5250$^g$&5.00&-1.26&s\\
4570$^h$& 1.10 & -2.42 & s&5070$^i$&4.69&-1.35&s\\
\hline\hline
\end{tabular}
\end{center}
Notes. $^a$p/s = photometric/spectroscopic determination.
$^b$\citet{gonbon09}
$^c$\citet{rammel05}
$^d$\citet{aam99}
$^e$\citet{black98}
$^f$NLTE analysis by \citet{mas08}
$^g$\citet{luck06}
$^h$\citet{mis01}
$^i$\citet{gehren06}
\end{table}


\section{Observations \label{sec:observations}}

The observations were collected at the CHARA Array \citep{tenbrummelaar05}, 
located at Mount Wilson Observatory (California), 
together with two beam combining instruments: 
CHARA Classic and FLUOR. 
CHARA Classic \citep{tenbrummelaar05} is a two-telescope, 
pupil-plane, open-air beam combiner 
working in both the $H$ and $K^{\prime}$ bands, and our observations 
correspond to the $K^{\prime}$ band 
(the central wavelength is 
$\lambda = 2.141\,\mu$m, from \citealt{bowsher10}). 
The raw data were reduced using the pipeline described in 
\citet{tenbrummelaar05}.
FLUOR \citep{coude98a,merand06} is a two-telescope beam combiner, but 
uses single-mode optical fibers for recombination.
Single-mode fibers efficiently reduce the perturbations induced by the 
turbulent atmosphere on the stellar light wavefront, 
as the injected light corresponds only to the mode guided by the fiber 
\citep{ruilier99,coude98b}. 
Most of the atmospherically corrupted part of the wavefront is lost into 
the cladding, and the beam combination therefore occurs between two 
almost coherent beams. 
This results in an improved stability of the measured fringe contrast. 
The FLUOR data 
reduction pipeline (\citealt{merand06}, 
see also \citealt{kervella04a}) is based on the Fourier 
algorithm and was developed by \citet{coude97}.

\begin{table}
\caption{CHARA observations of HD\,122563 and \gmb. }
\label{observations_log}
\begin{tabular}{lllllll}
\hline \hline
\noalign{\smallskip}
MJD & Inst. & $B$ & PA & $V$ & $\sigma(V)$ \\
(days) & & (m) & ($^{\circ}$) & &\\
\hline
HD\,122563\\
\noalign{\smallskip}
54603.427 & F  & 294.07 & -18.4 & 0.599 & 0.015 \\
54602.329 & F  & 284.90 & 8.3 & 0.648 & 0.015 \\
54602.309 & F  & 288.89 & 13.7 & 0.630 & 0.014 \\
54579.797 & C  & 312.54 & 240.4 & 0.562 & 0.067 \\
54579.784 & C  & 316.88 & 238.2 & 0.617 & 0.064 \\
54579.772 & C  & 320.35 & 236.5 & 0.534 & 0.063 \\
54579.760 & C  & 323.52 & 235.0 & 0.554 & 0.071 \\
54579.748 & C  & 326.10 & 233.6 & 0.543 & 0.101\\
54578.812 & C  &  308.35& 242.5 & 0.587 & 0.059\\
54578.801 & C  & 312.18 & 240.6 & 0.529 & 0.064\\
54578.786 & C  & 316.86 & 238.2 & 0.483 & 0.057\\
54578.771 & C  & 321.19 & 236.1 & 0.498 & 0.072\\
54578.755 & C  & 325.23 & 234.1 & 0.469 & 0.057\\
54645.698 & C  & 287.58 & 257.8 & 0.606 & 0.074 \\
54645.686 & C  & 290.38 & 254.8 & 0.584 & 0.080 \\
54645.676 & C  & 292.87 & 252.6 & 0.509 & 0.086 \\
54645.670 & C  & 294.78 & 251.1 & 0.527 & 0.075 \\
\hline
\gmb \\
\noalign{\smallskip}
54604.314 & F  & 330.49 & -11.5 & 0.744 & 0.023 \\
54604.279 & F  & 330.16 & -3.3 & 0.751 & 0.022 \\
54604.241 & F  & 330.23 & 5.7 & 0.756 & 0.022 \\
54459.022 & C  & 327.54 & 238.9 & 0.696 & 0.053 \\
54459.013 & C  & 326.45 & 237.4 & 0.724 & 0.047 \\
54459.005 & C  & 325.16 & 235.9 & 0.691 & 0.051 \\
54458.996 & C  & 323.67 & 234.4 & 0.733 & 0.031 \\
54458.959 & C  & 313.47 & 228.3 & 0.714 & 0.053 \\
54458.950 & C  & 310.34 & 227.1 & 0.734 & 0.055 \\
54458.935 & C  & 304.08 & 225.0 & 0.759 & 0.043 \\
54458.927 & C  & 299.91 & 223.9 & 0.757 & 0.036 \\
54421.060 & C  & 312.48 & 227.9 & 0.752 & 0.073 \\
54421.053 & C  & 310.12 & 227.0 & 0.756 & 0.064 \\
54421.047 & C  & 307.47 & 226.1 & 0.755 & 0.057 \\
54421.040 & C  & 304.69 & 225.2 & 0.759 & 0.064 \\
54421.032 & C  & 300.68 & 224.1 & 0.769 & 0.066 \\
54421.018 & C  & 293.59 & 222.4 & 0.776 & 0.056 \\
54421.009 & C  & 288.24 & 221.3 & 0.773 & 0.090 \\
\hline
\end{tabular}
\tablefoot{MJD is the average modified julian date of the exposures and 
Inst. the instrument code (F: FLUOR, C: Classic).}
\end{table}

We observed \object{HD\,122563} and 
\object{\gmb} 
in late 2007 and 2008 with FLUOR and Classic. 
The corresponding visibility measurements $V$ and uncertainties $\sigma(V)$
are listed in Table~\ref{observations_log} along with the 
projected baseline $B$ and the baseline position angle PA measured 
clockwise from North. 
To monitor the interferometric transfer 
function, we interspersed the observations of our two science targets 
with calibrator stars. 
The calibrators for the FLUOR observations were selected from the catalogue 
by \citet{merand05}, and 
these are listed in Table~\ref{calibrator_table}, 
and those for the CHARA Classic
observations used the calibrators presented in
Table~\ref{tab:Classic_calibrators}.

\begin{table}
\caption{FLUOR calibrator stars.}
\label{calibrator_table}
\begin{tabular}{llllll}
\hline \hline
\noalign{\smallskip}
Calibrator & Sp. type & $m_\mathrm{V}$ & $m_\mathrm{K}$ & UD (mas) & Target \\
\hline
\noalign{\smallskip}
\object{HD 129336} & G8III & 5.6 &3.4 & $0.98 \pm 0.01$ & HD 122563 \\
\object{HD 127227} & K5III & 7.5 & 4.0 & $0.84 \pm 0.01$ & HD 122563 \\
\object{HD 108123} & K0III & 6.0 &  3.7 & $0.93 \pm 0.01$ & \gmb\\
\object{HD 106184} & K5III & 7.7 & 3.5 & $0.98 \pm 0.01$ & \gmb \\
\hline
\end{tabular}
\end{table}

\begin{table}
\caption{CHARA Classic calibrator Stars}
\label{tab:Classic_calibrators}
\begin{tabular}{lccccc}
\hline \hline
\noalign{\smallskip}
Calibrator & Sp. type & $m_\mathrm{V}$ & $m_\mathrm{K}$ & UD (mas) &
Target \\
\hline
\noalign{\smallskip}
\object{HD 119550} & G2V & 6.9 & 5.3 & $0.389 \pm 0.027$ & HD 122563 \\
\object{HD 120066} & G0V & 6.3 & 4.9 & $0.479 \pm 0.033$ & HD 122563 \\
\object{HD 120934} & A1V & 6.1 & 6.0 & $0.198 \pm 0.014$ & HD 122563 \\
\object{HD 121560} & F6V & 6.2 & 4.8 & $0.460 \pm 0.030$ & HD 122563 \\
\object{HD 122365} & A2V & 6.0 & 5.7 & $0.238 \pm 0.016$ & HD 122563 \\
\object{HD 103799} & F6V & 6.6 &5.3 & $0.343 \pm 0.013$ & \gmb \\
\hline
\end{tabular}
\end{table}

We also retrieved archival observations of HD\,122563 in the $K$ band obtained with the Palomar Testbed Interferometer (PTI) \citep{colavita99a} between 1999 and 2002, and these are listed in 
Table~\ref{observations_log_PTI}. The data processing algorithm that was employed to reduce 
the PTI observations has been described in detail by \citet{colavita99b}. 
Due to the shorter baselines, the PTI observations resolve 
HD\,122563 marginally, and therefore do not strongly constrain its 
angular diameter. However, thanks to the relatively large number of observations, they provide 
an independent method for testing any bias 
in the CHARA observations.

\section{From visibilities to limb-darkened angular diameters\label{sec:diameters}}

We employed a non-linear, least-squares fitting routine in IDL (MPFIT,
\citealt{markwardt09}) to fit 
uniform disk and limb-darkened visibility functions for a single star
to 
the calibrated data points
\citep[see][]{hanbury74, boyajian12}. 
We obtained a uniform disk diameter for HD\,122563 and \gmb\ of 
$\theta_{\rm UD} = 0.924 \pm 0.011$ mas and 
$\theta_{\rm UD} = 0.664 \pm 0.015$ mas, respectively.
We used the linear limb-darkening
coefficients from \citet{claret00} assuming 
a 
[Fe/H]$=-2.5$, $T_{\rm eff} = 4500$ K, and $\log g = 1.0$ for HD~122563 
and [Fe/H]$=-1.5$, $T_{\rm eff} = 5000$ K, and $\log g = 4.5$ for \gmb.
The assumptions
on these parameters on the adopted coefficients have minimal influence
on the final limb-darkened diameter, adding uncertainties of only a few
tenths of a percent, well within the errors of our diameter measurements.
We obtained 
$\theta_{\rm LD} = 0.948 \pm 0.012$ and 
$\theta_{\rm LD} = 0.679 \pm 0.015$ for HD\,122563 and \gmb, respectively.

\begin{figure} %
\includegraphics[width=0.5\textwidth]{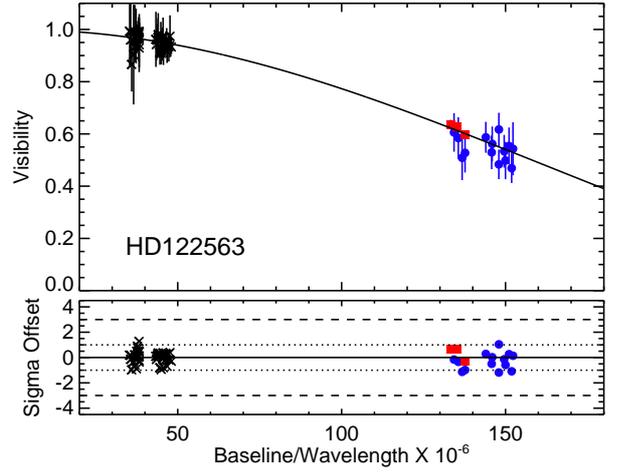}
\caption{Calibrated observations for PTI ({\it black crosses}), CHARA
Classic ({\it blue circles}) and CHARA FLUOR ({\it red squares}) data
plotted with the 1D limb-darkened visibility function fit for HD~122563. See
Section~\ref{sec:diameters} for details.}
\label{fig:HD122563_diameter}
\end{figure}

\begin{figure} %
\includegraphics[width=0.5\textwidth]{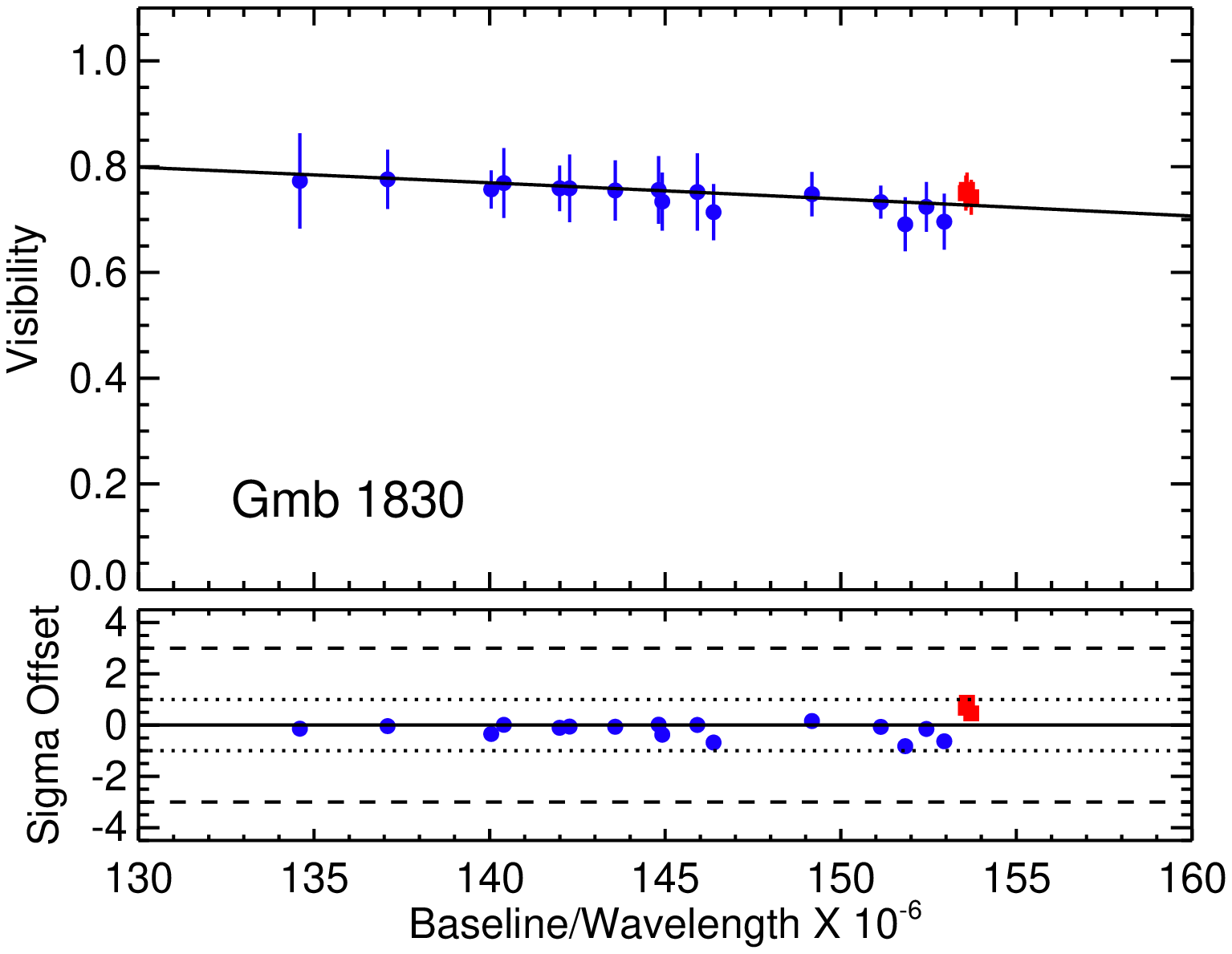}
\caption{Calibrated observations for CHARA Classic ({\it blue
circles}) and CHARA FLUOR ({\it red squares}) data plotted with the
limb-darkened visibility function fit for \gmb. See
Section~\ref{sec:diameters} for details.}
\label{fig:HD103095_diameter}
\end{figure}

We obtained a reduced $\chi^2$ value of 0.28 for HD\,122563
and 0.18 for \gmb\ from the fits.  
These values, much less than 1, are indicative of our individual measurement
errors being over estimated.
We show the data 
and the visibility function fits for HD~122563 and \gmb\ in
Figures~\ref{fig:HD122563_diameter} and \ref{fig:HD103095_diameter}.
The results from the fits to the data yield
1D limb-darkened angular diameters of HD122563 and \gmb\ with 
a precision of $\sim$2\% (see Table~\ref{tab:diameters}),
respectively.




\begin{table}
\caption{PTI observations of HD\,122563.}
\label{observations_log_PTI}
\begin{tabular}{lllll}
\hline \hline
\noalign{\smallskip}
MJD & $B$ &  $V$ & $\sigma(V)$ \\
(days) & (m) \\
\hline
\noalign{\smallskip}
51255.371  & 105.63 & 0.914 & 0.043 \\
51255.384  & 104.32 & 0.940 & 0.030 \\
52000.309  & 107.55 & 0.939 & 0.017 \\
52000.320  & 106.67 & 0.947 & 0.015 \\
52000.329  & 105.88 & 0.948 & 0.016 \\
52000.334  & 105.37 & 0.948 & 0.018 \\
52023.231  & 108.58 & 0.933 & 0.028 \\
52023.271  & 105.38 & 0.957 & 0.048 \\
52023.313  & 101.08 & 0.958 & 0.031 \\
52041.196  & 107.66 & 0.974 & 0.080 \\
52041.206  & 106.83 & 0.933 & 0.056 \\
52041.229  & 104.73 & 0.957 & 0.035 \\
52041.236  & 103.99 & 0.954 & 0.031 \\
52041.258  & 101.62 & 0.908 & 0.043 \\
52041.267  & 100.70 & 0.912 & 0.047 \\
52044.236  & 103.14 & 0.910 & 0.046 \\
52044.243  & 102.39 & 0.948 & 0.060 \\
52044.276  & 99.00 & 0.959 & 0.061 \\
52044.283  & 98.36 & 0.943 & 0.061 \\
52306.520  & 102.86 & 0.973 & 0.072 \\
52306.522  & 102.68 & 0.941 & 0.061 \\
52306.535  & 101.36 & 0.963 & 0.066 \\
52306.552  & 99.54 & 0.980 & 0.083 \\
52306.554  & 99.37 & 0.969 & 0.074 \\
52306.569  & 97.96 & 0.962 & 0.106 \\
52328.424  & 84.82 & 0.981 & 0.035 \\
52328.432  & 85.50 & 0.993 & 0.040 \\
52328.450  & 86.41 & 0.968 & 0.047 \\
52328.459  & 86.47 & 0.941 & 0.087 \\
52329.419  & 84.57 & 0.965 & 0.056 \\
52329.426  & 85.24 & 0.984 & 0.023 \\
52329.438  & 86.08 & 0.973 & 0.022 \\
52329.445  & 86.36 & 0.966 & 0.030 \\
52329.464  & 86.33 & 0.986 & 0.019 \\
52329.471  & 86.04 & 0.979 & 0.031 \\
52329.490  & 84.51 & 0.938 & 0.030 \\
52329.498  & 83.52 & 0.933 & 0.054 \\
52329.516  & 80.61 & 0.960 & 0.052 \\
52353.369  & 85.86 & 0.970 & 0.053 \\
52353.385  & 86.46 & 0.988 & 0.056 \\
52353.392  & 86.46 & 0.993 & 0.067 \\
52353.426  & 84.27 & 0.975 & 0.105 \\
52353.433  & 83.39 & 0.987 & 0.107 \\
52353.451  & 80.54 & 0.989 & 0.119 \\
52353.455  & 79.88 & 0.993 & 0.139 \\
52359.378  & 86.42 & 0.929 & 0.093 \\
52359.386  & 86.18 & 0.963 & 0.084 \\
52359.423  & 82.58 & 0.904 & 0.191 \\
52359.431  & 81.24 & 0.866 & 0.104 \\
\hline
\end{tabular}
\end{table}

\subsection{3D limb-darkened angular diameter for HD\,122563}

Convection plays a very important role in the determination of stellar limb-darkening. 
It has been shown that a 3D hydrodynamical treatment of the surface layers can lead to a significative change of temperature gradients  compared to 1D hydrostatic modeling, which consequently affects
the center-to-limb intensity variation \citep[e.g.][]{2002ApJ...567..544A,bigot06,2009A&A...508.1403P,2010A&A...524A..93C,2011A&A...534L...3B,2012A&A...539A.102H}. The 3D/1D limb-darkening correction for a giant star can be very significant  \citep[see Fig.~6 of ][]{2010A&A...524A..93C} and is generally much stronger than for  a dwarf star.
We therefore used a radiative-hydrodynamical (RHD) surface convection simulation of a red giant for HD\,122563 to determine the 3D limb-darkened angular diameter. 
The parameters of the model are 
$\langle T_{\rm{eff}} \rangle =4627\pm14$K 
(temporal average and standard deviation of the effective 
temperature), [Fe/H]=$-$3.0, and $\log g$ = 1.6 
\citep{collet09,2010A&A...524A..93C}. The computational domain of the RHD simulation represents only a small portion of the stellar surface ($\sim$1/30 of the circumspherence), however, it is sufficiently large to contain enough granules ($\sim$10-15) at each time step of the simulation. Hydrodynamical equations are solved on a staggered mesh with a conservative scheme. Details of the computation can be found  in \citet{collet09}.


\begin{table*}

\caption{Angular diameters}
\label{tab:diameters}
\begin{tabular}{lccccc}
\hline \hline
\noalign{\smallskip}
& \# of &  $\theta_{\rm UD} \pm \sigma$ & $\theta_{\rm
1D} \pm \sigma$ & $\theta_{\rm 3D} \pm \sigma$\\
Star & Observations &  (mas) & (mas) &
(mas) \\
\hline
\noalign{\smallskip}
\gmb\ & 18 &  $0.664 \pm 0.015$ &
$0.679 \pm 0.015$ & \dotfill \\
HD~122563  & 66 &  $0.924 \pm 0.011$ &
$0.948 \pm 0.012$ & $0.940 \pm 0.011$ \\
\hline
\end{tabular}
\end{table*}

We computed emergent intensity for a representative series of simulated snapshots and  for wavelengths corresponding 
to the FLUOR filter ($2.14\pm0.26$ $\mu$m, equivalent to that for CHARA) using the 3D pure-LTE\footnote{Pure-LTE refers to when 
the source function is equal to
the Planck function} radiative transfer code {\sc Optim3D}
\citep{chiavassa09}.
It considers the
Doppler shifts due to convective motions. 
Radiative transfer is solved monochromatically using 
pre-tabulated extinction coefficients for the same chemical 
compositions as the RHD simulations. 
It also uses
the same extensive atomic and molecular opacity data 
as the latest generation of
MARCS models \citep{gustafsson}.

\begin{figure}
  \centering
 \includegraphics[width=0.50\textwidth]{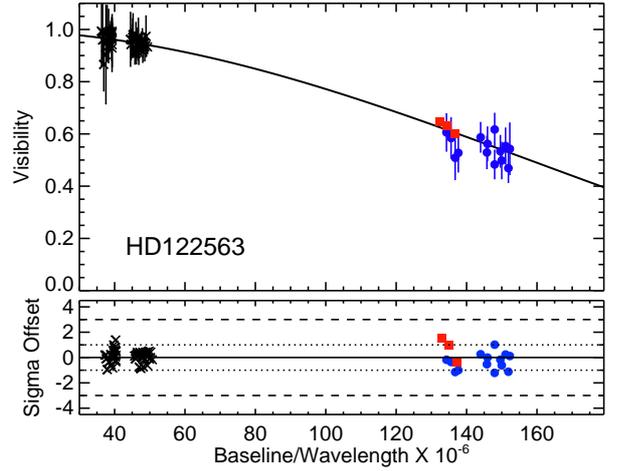} 
     \caption{
Best matching 3D-RHD synthetic visibility curves and 
PTI ({\it black crosses}), Classic ({\it blue circles}), 
and FLUOR ({\it red squares}) data for HD\,122563. 
       \label{fig:3D}}
  \end{figure}

For each time-step, we solve the radiative transfer equation for different 
inclinations with respect to the vertical whose cosines are 
$\mu{\equiv}$[1.000, 0.989, 0.978, 0.946, 
0.913, 0.861, 0.809, 0.739, 0.669, 0.584, 0.500, 0.404, 0.309, 0.206, 0.104].
 From these limb-darkened intensities, we derived the monochromatic visibility curves using the Hankel Transform. The visibilities are then averaged with the transmission function of the instrument in the considered filter wavelength domain.  The procedure used in this work is the same as that of 
\citet{2011A&A...534L...3B}. The synthetic visibilities are used to fit the
interferometric $K$ band observations given in Tables 
\ref{observations_log} and \ref{observations_log_PTI}.

Figure~\ref{fig:3D} displays the best fit of the visibility 
curve to the data 
that results 
in an angular diameter of $\theta_{\rm{3D}}$=0.940 $ \pm $ 0.011 mas 
(Table~\ref{tab:diameters}) with a $\chi^2 = 0.35$. 
Its value lies between that of the uniformed disk and 1D limb-darkened diameters. This is a consequence of the fact that in realistic 3D hydrodynamical treatment of the stellar surface, the emergent intensity is less limb-darkened than the 1D hydrostatic case. We note that the choice of the exact fundamental parameters  of the 3D simulation does not influence the limb-darkened intensity and the derived angular diameter by much.

The 3D/1D correction is important for determining the zero point
of the effective temperature scale:
\citet{2010A&A...524A..93C} (Table~3) showed that, 
in the case of metal-poor stars like the one analyzed in this work, 
$\theta_{\rm{3D}}/\theta_{\rm{1D}}\sim2\%$ in the $K$ band (3.5$\%$ 
in the visible).
This can result in corrections to the effective temperature of $\sim$40 K
in the $K$ band.
In this case the resulting correction to the effective temperature is $\sim$15 K
(see Sect.~\ref{sec:fparamsobs}).

We note that \citet{gonbon09} 
predict 1D limb-darkened angular diameters for both
of these stars using the infra-red flux method (IRFM)\footnote{The IRFM 
allows one to calculate 
\teff\ by comparing the ratio of infra-red to
bolometric flux observed from Earth to the true intrinsic value obtained from
theoretical models (see e.g. \citealt{cas08})}.
For HD\,122563 they predict $\theta_{\rm 1D}$ = 0.84 \pmm\ 0.04 mas and 
for \gmb\ $\theta_{\rm 1D}$ = 0.61 \pmm\ 0.02 mas; both values are 
lower than our derived values.  For HD\,122563 
this could be due to the fact that
they use 2MASS photometry that is saturated for this star (saturation limit 
$K_{\rm s} \sim 4.0$, \citealt{cut03}).


\section{Constraints on stellar evolutionary models \label{sec:fparams}}


\subsection{Observed parameters\label{sec:fparamsobs}}
The list of observed parameters are summarized in the top part of
Table~\ref{tab:observ}.  
The magnitudes in the $V$ band are taken from 
\citet{jp11}, those in the $K$ bands are from 
\citet{duc02} for HD\,122563 and \citet{cut03} for \gmb, and 
the Hipparcos parallax from \citet{hipparcos07}.
For HD\,122563, we estimate an interstellar extinction of 
A$_V$ = 0.01 mag 
based on its galactic coordinates and distance.
The bolometric flux $F_{\rm bol}$ is obtained by combining $m_V$, A$_V$, and 
the bolometric correction
BC$_V$, 
where BC$_V$ is obtained by
interpolating the tables for giant stars 
from \citet{alonso99}.
We started with an initial \teff\ of 4530\,K (and [Fe/H] = -2.5) 
to calculate BC$_V$ from the tables, and then used this value to determine
an initial F$_{\rm bol}$.  
Using the initial F$_{\rm bol}$ and the derived $\theta_{\rm LD}$ we 
determined \teff\ (see below).
This new \teff\ was then used to rederive BC$_V$, $F_{\rm bol}$ and \teff, 
and we iterated until we converged on the final \teff\ of 
4582\,K using BC$_{V} = -0.472$ for $\theta_{\rm 1D}$ and 
4598\,K using BC$_{V} = -0.466$ for $\theta_{\rm 3D}$.
We note that adopting these \teff\ and interpolating the tables 
from \citet{houdas00} yields BC$_V$ within our error bars
(BC$_V \sim -0.46$). 
For \gmb, we used F$_{\rm bol}$ and A$_V$ 
from \citet{boyajian12}, and then indirectly calculated BC$_V$.
We do not subsequently use BC$_V$ in this work but we report the value for 
reference.
Both the 1D and 3D limb-darkened angular diameters $\theta_{\rm LD}$ 
are given for HD\,122563 and the 1D diameter is given for \gmb\
(see Table~\ref{tab:diameters}).
The {\it surface brightness} relations from \citet{kervella04c,kervella04b} 
were used
to provide an estimate of the 1D 
limb-darkened angular diameter $\theta_{\rm pred}$.
The predicted values are lower than the derived values, although 
for HD\,122563 the agreement is quite good ($\theta_{\rm pred} = 0.928$ mas, 
$\theta_{\rm 1D} = 0.948$ mas).
These relations have been calibrated with a large sample of stars.
However, the 
obvious lack of reliable measurements of metal-poor stars may lead to slight
biases in the angular diameters predicted using these methods.

Combining the above mentioned measurements we determined the 
{\it observed} or {\it model-independent} fundamental properties of both stars;
absolute magnitude M$_V$, \teff, $L$, and $R$, where  
\teff\ is derived using the equation 
$T_{\rm eff} = \left ( \frac{4}{\sigma_{\rm SB}} \frac{F_{\rm bol}}{\theta^2} \right)^{0.25}$,
$\sigma_{\rm SB}$ is the Stephan-Boltzmann constant and $\theta$ is the 
limb-darkened angular diameter.

The most recently published NLTE spectroscopic analysis of HD\,122563
yielded [Fe/H] = $-2.53 \pm 0.02$ dex (see Table~\ref{tab:teffs}, \citealt{mas08}).
The same authors also derived NLTE abundances for two $\alpha$ elements:
[Mg/H] = $-2.2$ and [Ca/H] = $-2.3$ to $-2.4$ dex.
In the PASTEL catalogue there are 15 spectroscopic determinations of 
[Fe/H] since 1990 (mostly LTE) with a mean value of $-2.7$ dex, or 
5 determinations since 2000 with a mean of $-2.6$.
The mean metallicity, which is a mixture of Fe peak and $\alpha$
elements then becomes [$Z/X$]$_s$ = --2.3 \pmm\ 0.1 dex.
Spectroscopic \logg\ values typically vary between 1.1 and 1.5 dex (see
Table~\ref{tab:teffs}).

For \gmb, \citet{gehren06} derived [Fe/H] = $-1.35 \pm 0.10$ dex from 
Fe II lines,
which are not supposed to be affected by NLTE, and 
an NLTE [Mg/Fe] abundance of +0.3.
Using the latter for all $\alpha$ elements, this implies a 
[$Z/X$]$_{\rm s} = -1.3 \pm 0.1$ dex.
The spectroscopic \logg\ of this star has been estimated to be
$\sim$4.70 \citep{thev99} 
from NLTE studies but using a 
temperature hotter by about 200 K.  
The \teff\ reported in this work would result in a downward 
revision of this number.

\begin{table*}
\caption{Observed parameters of HD\,122563 and \gmb.\label{tab:observ}}
\begin{tabular}{lcccc}
\hline\hline\\
Observation & HD\,122563 && \gmb\\
& 1D & 3D\\
\hline
m$_V$ (mag) & 6.19 \pmm\ 0.02&&6.45 \pmm\ 0.02\\
m$_K$ (mag) & 3.69 \pmm\ 0.04 && 4.37 \pmm\ 0.03\\
$\pi$ (mas) & 4.22 \pmm\ 0.35 && 109.99 \pmm\ 0.41\\
$[$Z/X$]_s$ (dex) & -2.3 \pmm\ 0.1 && -1.3 \pmm\ 0.1\\
BC$_V$ (mag) & -0.472 \pmm\ 0.02 &-0.466 \pmm\ 0.02 & -0.23 \pmm\ 0.01$^{a}$ \\
A$_V$ (mag) & 0.01 \pmm\ 0.01 && 0.00 \pmm\ 0.01\\
$F_{\rm bol}$ (erg s$^{-1} {\rm cm}^{-2} \times 10^{-8}$)&
13.23 \pmm\ 0.37$^b$&
13.16 \pmm\ 0.36$^b$ 
&8.27 \pmm\ 0.08\\
$\theta_{\rm pred}^c$ (mas) & 0.928 \pmm\ 0.019 & &0.630 \pmm\ 0.013 \\
$\theta_{\rm LD}$ (mas) &0.948 \pmm\ 0.012 &0.940 \pmm\ 0.011$^d$ & 0.679 \pmm\ 0.015\\
\\
M$_V$ (mag) &-0.69 \pmm\ 0.03 &-0.69 \pmm\ 0.03&6.66 \pmm\ 0.02 \\
$T_{\rm eff}$ (K) & 4585 \pmm\ 43 &4598 \pmm\ 41 & 4818 \pmm\ 54 \\
$L$ (L$_{\odot}$) & 232 \pmm\ 6 &230 \pmm\ 6$^e$ & 0.213 \pmm\ 0.002$^e$ \\
$R$ (R$_{\odot}$) & 24.1 \pmm\ 1.9 &23.9 \pmm\ 1.9 & 0.664 \pmm\ 0.015\\
\hline\hline\\
\end{tabular}

Notes. $^a$BC$_V$ is derived assuming F$_{\rm bol}$ from \citet{boyajian12}.
$^bF_{\rm bol}$ is derived using m$_V$, A$_V$, and BC$_V$ from \citet{alonso99}.
$^c\theta_{\rm pred}$ is the predicted angular diameter using the surface-brightness 
relations from \citet{kervella04c,kervella04b}.
$^d$3D limb-darkened angular diameter.
$^eL$ calculated from $F_{\rm bol}$ and $\pi$.
\end{table*}

\subsection{CESAM2k models\label{sec:models}}

In order to interpret the observations of HD\,122563 and \gmb\ we used
the CESAM2k stellar evolution and structure code \citep{mor97,ml08}.
We tested the models using three different equations of state (EOS):
the classical EFF EOS \citep{eff} with/without 
Couloumb corrections (CEFF/EFF), 
and 
the OPAL EOS \citep{opaleos}, and we found small differences
in the derived parameters for \gmb\ only.  
For all of the models we used the OPAL opacities \citep{ri92} 
supplemented with \citet{af94} molecular opacities.
The p-p chain, CNO, and triple-$\alpha$ nuclear reactions were calculated
using the NACRE rates \citep{ang99}.
We adopted the solar abundances of \citet{gn93} ($Z_{\odot} = 0.017$, 
$X_{\odot} = 0.694$) and used the MARCS model atmospheres \citet{marcs03}.
Microscopic diffusion was taken into account for \gmb\ 
and follows the treatment described
by \citet{bur69}, and we introduced 
extra mixing by employing a parameter, Re$\nu=1$, as prescribed
by \citet{morelthev02} in order to slow down the depletion of helium and 
heavy elements.
  For HD\,122563 no observable difference is found between
diffusion and non-diffusion models for giants, except a small effect on the age
of the star, i.e., for the same parameters the model with diffusion fits the
observational data with an
age $\sim$0.3 Gyr older than
the non-diffusion model. 
Convection in the outer envelope 
is treated by using the  mixing-length theory described by \citet{egg72},
where $l = \alpha H_p$ is the mixing-length that tends to 0 as the 
radiative/convective borders are reached, $H_p$ is the pressure scale
height,  and 
$\alpha$ is an adjustable parameter.
To match the solar
luminosity, \teff, and oscillation frequencies (while including diffusion)
we find a value of $\alpha = 2.04$.
We note that we did not include convective overshooting in our models because
the primary effect that this extra parameter has
on the determination of the stellar
model is the age.
This means that it is possible to find two equivalent
stellar models with the same stellar parameters 
that differ only by age and the value of the overshoot parameter. 
Since we have no observable constraint to distinguish between these two
parameters we chose not to include it.

Each stellar model is defined by a set of input model parameters ---
mass $M$, initial helium content $Y_{\rm i}$, 
initial metal to hydrogen ratio \zx,
age $t$, and the mixing-length parameter $\alpha$ ---
and these result in model observables, such as a model \teff\ and a model
$L$.
By varying the parameters $M$, $Y_{\rm i}$, \zx, $t$, and $\alpha$ we 
aimed to find models that fitted the luminosity, 
\teff, and metallicity constraints as outlined in Table~\ref{tab:observ}.
We stopped the evolution of the models when an age of 14 Gyr was reached.
For HD\,122563 we chose to use the constraints from the more realistic 
3D models, although we note that the difference between the 1D and 3D 
constraints leads to only very slight changes in the 
parameters of the stellar models (see Sect.~\ref{sec:giantmodels} below).

\subsection{Stellar parameters}

Figure~\ref{fig:gmbmodel} shows two HR diagrams with the 
observational error boxes of both stars (left/right = HD\,122563/\gmb) as well 
some models that lie somewhat away from the central
position of the box, illustrative of the uncertainties that
we find in the stellar parameters (see below).
Table~\ref{tab:params} 
lists the stellar parameters for both stars using the classical
EFF, and for \gmb\ we also give the stellar properties for the CEFF and OPAL
EOS models.

Given the few independent observational 
constraints and the large number
of adjustable parameters in the models, a classical error analysis is not
possible for both stars.  
In order to estimate the uncorrelated uncertainties we
changed each of the reference parameters of the models individually
until we reached the edges of the error box in the HR diagram, or
the limits of each parameter, e.g. we did not test $Y_{\rm i} < 0.20$.
These are the uncertainties that are given in the top part of
Table~\ref{tab:params}.
For the uncertainty in the age we give the $1\sigma$ uncertainty which
corresponds to the central models approaching the upper and lower limit in
luminosity (first number) and we also give 
the range of possible ages 
while considering the uncertainties in the 
four model parameters (second uncertainty).
We also list the model observables and their uncertainties in the lower part
of the table.
We note that the uncertainties in the model observables cover the full range
of values while considering the individual changes in each of the 
four model parameters.

\subsubsection{HD\,122563\label{sec:giantmodels}}

For HD\,122563 using the EFF EOS description 
we found a best model with $M =  0.855$ \msol, 
$Y_{\rm i} = 0.245$, $\alpha = 1.31$, 
and $t = 12.6$ Gyr.  
We fixed $Z_{\rm i}/X_{\rm i} = 0.0001$ in order
to have the correct observed [$Z/X$]$_s$. 
This model is illustrated in Figure~\ref{fig:gmbmodel} 
(left panel) by the thick line and is clearly labelled.
We also show some models which illustrate the reported uncertainties:
the black continuous line shows a model when $\alpha$ is changed by
0.08, the red continuous line shows the central model when $Y_{\rm i}$ is
decreased by 1$\sigma$, and finally the red dashed-dotted line shows the effect
of increasing the mass by 1$\sigma$.  We note that if we increase the mass to 
more than 0.88 \msol\ then the age of the model becomes too small ($<10$ Gyr)
if we 
are to consider the giant a halo star.
We also found correlations among the parameters $M$, $Y_{\rm i}$, and $\alpha$, 
and adjusting two of the three at a time by a small amount reproduces 
the position of the central model, e.g.
if we fix $M$ then $\Delta Y = +0.01 <=> \Delta \alpha = -0.01$.  
However, these correlations are adequately
accounted for in the uncertainties.

The dotted error box shows the constraints if we consider the 1D 
limb-darkened angular diameter.  
The stellar parameters of the model that passes through the center of the 
box need small adjustments when compared to the 3D diameter constraints.
In particular, decreasing either $M$ or $Y_i$ alone by less than 
1$\sigma$ or decreasing $\alpha$ by $\sim 0.03$ (or a combination of the three)
would reproduce the central position of the error box with a slightly
higher age.  If the temperature constraint were even 
lower, then the only viable option would be reducing the mixing-length parameter
$\alpha$, because reducing $M$ or $Y_i$ by much more would result in a model
that fails to reach the minimum luminosity before 14 Gyr.

Inspecting the stellar parameters in Table~\ref{tab:params} we 
highlight the excellent precision obtained in the mass of this single star.
Generally such precisions can only be obtained if the star
is in a binary system, where the solutions are then model-independent.
Combining this value with the well-determined radius yields a very 
precisely determined \logg\ (= 1.60 \pmm\ 0.04 dex).
This value is 
larger than most values used for spectroscopic analyses which 
typically ranges from 1.1 - 1.5 dex (see Table~\ref{tab:teffs}).
More recent work using 3D hydro-dynamical simulations for stellar
atmospheres quote values of 1.1 - 1.6 dex 
(see e.g. \citealt{barbuy03}, \citealt{collet09}, \citealt{ram10}).

\subsubsection{\gmb\label{sec:dwarfmodels}}

\begin{figure*}
\includegraphics[width=0.48\textwidth]{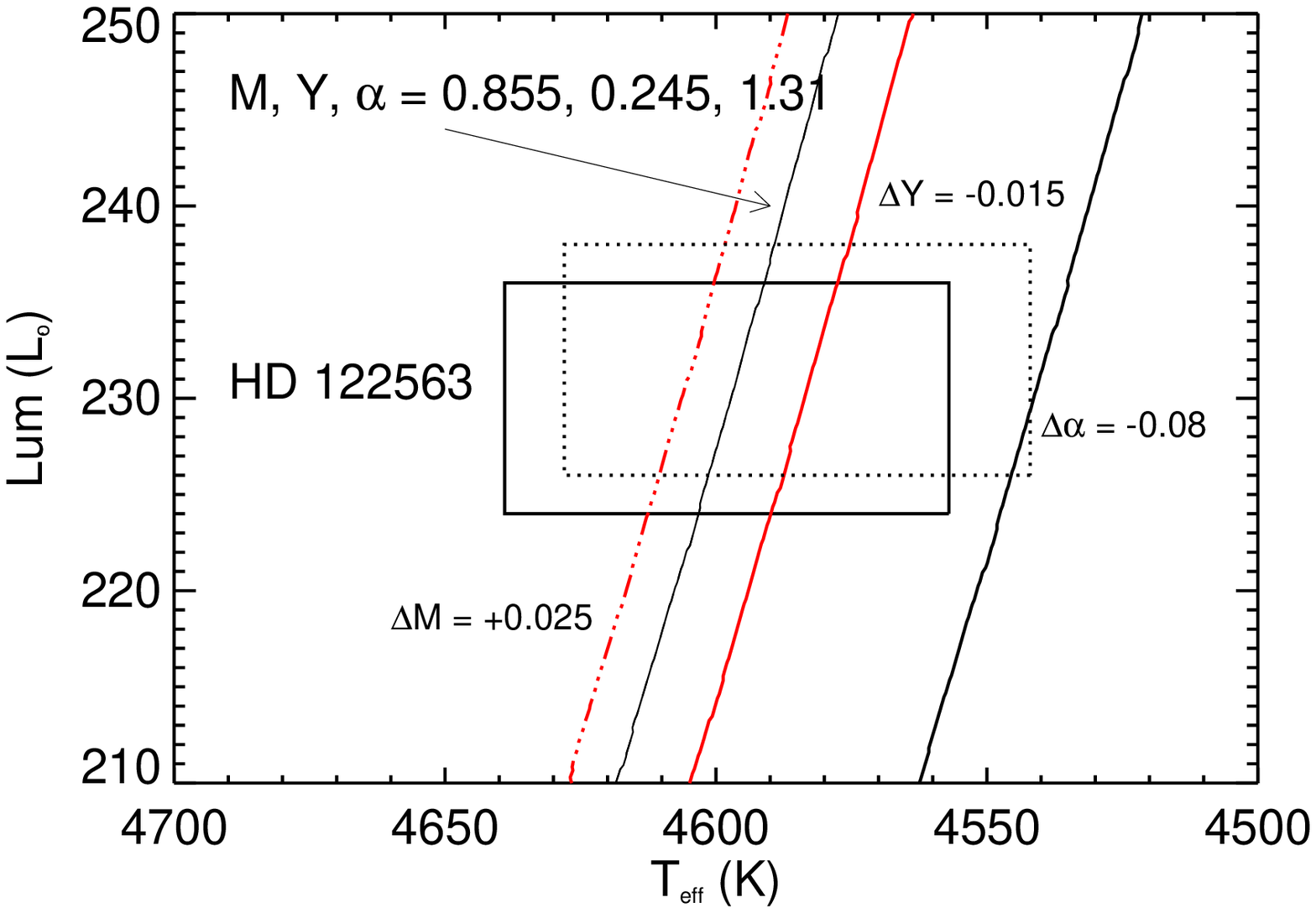}
\includegraphics[width=0.48\textwidth]{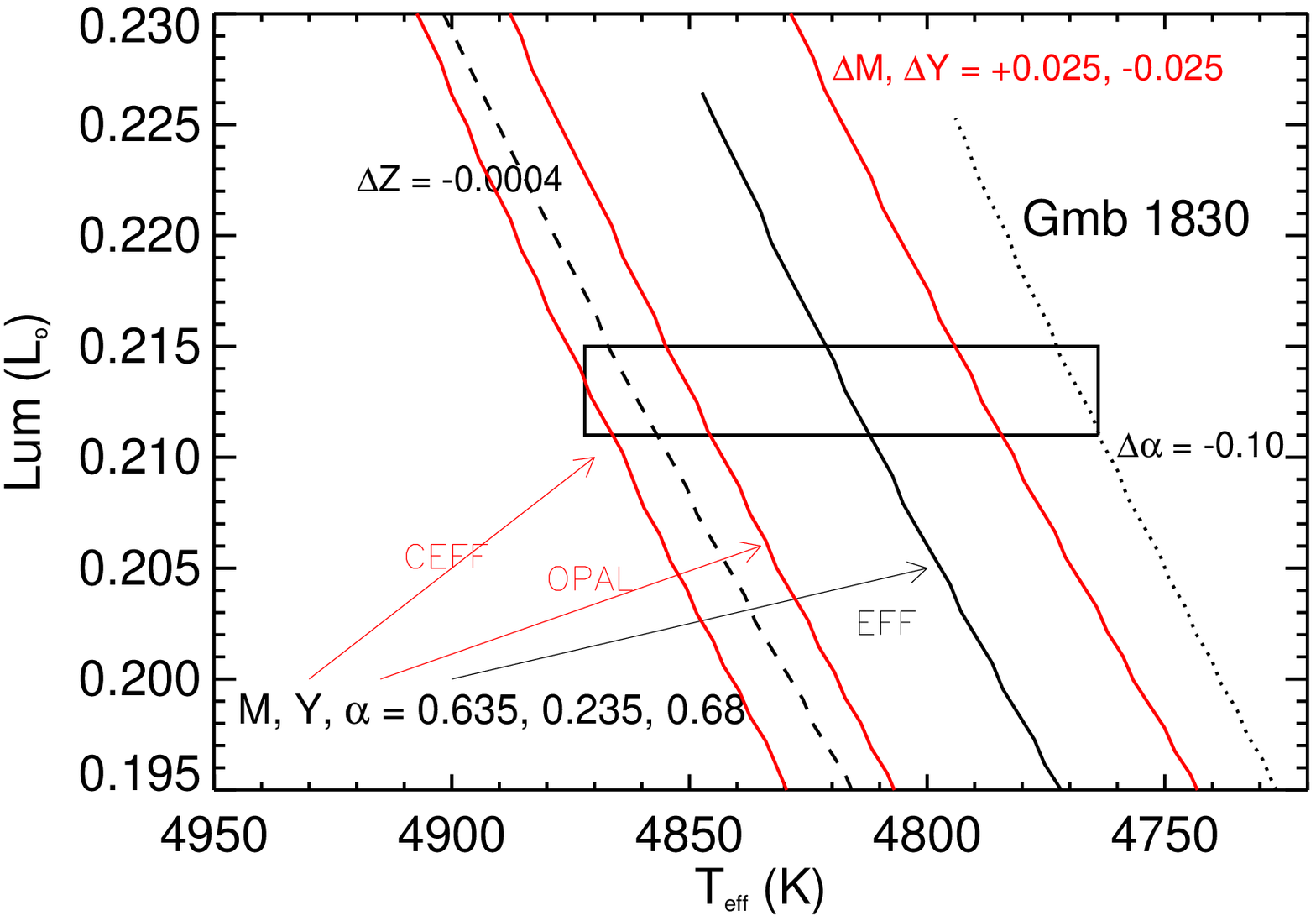}
\caption{HR diagram showing the observational error boxes for 
HD\,122563 (left) and \gmb\ (right). 
Both figures show
stellar models that pass through the error boxes which allow us to 
determine the stellar model properties and their uncertainties.
Each panel shows the adopted central models (with arrows) obtained by
considering the HR and metallicity constraints.
Other models are also indicated to highlight the 
parameter uncertainties and correlations.
Refer to Sects.~\ref{sec:giantmodels} and \ref{sec:dwarfmodels} for details.
{\label{fig:gmbmodel}}}
\end{figure*}

In Table~\ref{tab:params} right three columns we summarize the stellar
parameters for \gmb\ using the EFF, CEFF, and OPAL EOS.  
In Figure~\ref{fig:gmbmodel} we show the central model for the EFF EOS
(arrow with 'EFF') 
with illustrative uncertainties.   
The model parameters are M = 0.635 \pmm\ 0.025 \msol, $Y_{\rm i}$ = 0.235 
\pmm\ 0.025, [$Z/X$] = 0.0016 \pmm\ 0.0004, $\alpha = 0.68 \pm 0.10$, 
and $t = 12.0 \pm 0.2^{+1.8}_{-2.2}$.
We also show a CEFF and OPAL EOS evolution track using the central parameters
obtained with the EFF model.  
A qualitative difference between the three EOS is notable,
however, considering the uncertainties in the stellar parameters, these 
differences are not significant.

The uncertainties reported in Table~\ref{tab:params} 
do not consider all of the correlations among the parameters.
For example, reducing the mass by 1$\sigma$ implies a necessary increase in 
$Y_{\rm i}$ by $1\sigma$ in order to remain
inside the error box and vice versa.
In Figure~\ref{fig:gmbmodel} we show effects of the uncertainties on the
central model; the dotted black line shows the effect of decreasing
$\alpha$ by 0.10, the dashed black line shows the effect of decreasing 
$Z_{\rm i}/X_{\rm i}$ by $1\sigma$ (denoted by $\Delta Z$ in figure), and 
the red continuous line right of the central model is when 
the mass is decreased by 1$\sigma$ and $Y_{\rm i}$ increased by 1$\sigma$.
We note that by decreasing/increasing the mass or $Y_{\rm i}$ alone leads
to a very young stellar model (not consistent with a halo star), a model
that is too hot, or at the age of 14 Gyr the luminosity does not reach the 
minimum required 0.210 \lsol.

\begin{table*}
\caption{Stellar properties and 1$\sigma$ uncertainties 
derived from modelling HD\,122563 
(no atomic diffusion) and \gmb\ (with atomic diffusion).
\label{tab:params}}
\begin{tabular}{lcccccccc}
\hline\hline\\
& HD\,122563 & \gmb\\
Property& \multicolumn{2}{c}{EFF EOS} & CEFF EOS& OPAL EOS\\
\hline
$M$ (M$_{\odot}$)&0.855 \pmm\ 0.025 & 0.635 \pmm\ 0.025 & 
0.625 \pmm\ 0.015 & 0.620 \pmm\ 0.020\\
$Y_{\rm i}$ &0.245 \pmm\ 0.015& 0.235 \pmm\ 0.025 &
0.230 \pmm\ 0.020 & 0.235 \pmm\ 0.025\\
$Z_{\rm i}$/X$_{\rm i}$ &0.00010 \pmm\ 0.00002 & 0.0016 \pmm\ 0.0004 &
0.0016 \pmm\ 0.0004& 0.0016 \pmm\ 0.0004\\
$\alpha$ &1.31 \pmm\ 0.08 &0.68 \pmm\ 0.10 &
0.63 \pmm\ 0.08 & 0.65 \pmm\ 0.10 \\
Age (Gyr) & 12.6 \pmm\ 0.1$^{+1.0}_{-1.5}$ & 12.1 \pmm\ 0.2$^{+1.8}_{-2.2}$ &
12.7 \pmm\ 0.3$^{+1.3}_{-2.1}$ & 12.3 \pmm\ 0.3$^{+1.3}_{-2.3}$\\
\\
$R$ (R$_{\odot}$)&24.1 \pmm\ 1.1& 0.665 \pmm\ 0.014 
&0.665 \pmm\ 0.015 &0.665 \pmm\ 0.015 \\
$L$ (L$_{\odot}$)&230 \pmm\ 7& 0.213 \pmm\ 0.002 & 0.213 \pmm\ 0.002
& 0.213 \pmm\ 0.002 \\
\teff\ (K) & 4598 \pmm\ 42 & 4815 \pmm\ 52 
& 4814 \pmm\ 53 & 4815 \pmm\ 50\\
\logg\ (dex) &1.60 \pmm\ 0.04 & 4.60 \pmm\ 0.02 
&4.59 \pmm\ 0.02 & 4.58 \pmm\ 0.02\\
$[$Z/X$]_s$ &-2.38 \pmm\ 0.10& -1.32 \pmm\ 0.11
&-1.32 \pmm\ 0.11 & -1.33 \pmm\ 0.11\\
\mlsep$_{\rm pred}^a$ ($\mu$Hz)&1.06 \pmm\ 0.06 & 198 \pmm\ 6 
&197 \pmm\ 7 & 196 \pmm\ 6\\
\numax$_{\rm pred}^a$ ($\mu$Hz) &5.16 \pmm\ 0.38 & 4886 \pmm\ 190
&4809 \pmm\ 199 & 4768 \pmm\ 188\\
\hline\hline\\
\end{tabular}

Notes: The first five values are the input parameters of the model and 
the other values are properties of these models.
The uncertainties are derived by perturbing each of the model parameters 
individually until the edge of the error box is reached.\\
$^a$\mlsep\ and \numax\ are the predicted seismic quantities
according to the scaling relations 
from \citet{bro94,kje95}, and the range of values listed consider 
all the uncertainties in the five model parameters.
\end{table*}

\subsection{Asteroseismic Constraints}
In Table~\ref{tab:params} we predict two global asteroseismic quantities 
\mlsep\ and \numax\ based on scaling relations 
\citep{bro94,kje95} corresponding to the reference models.  
Both quantities are proportional to the mass and radius of the star, with 
the latter also having a small \teff-dependence;
\begin{equation}
\hspace{0.5cm}
\frac{\langle \Delta \nu \rangle}{\langle \Delta \nu \rangle_{\odot}} 
\approx M^{0.5}R^{-1.5},  \hspace{0.8cm}
\frac{\nu_{\rm max}}{\nu_{\rm max, \odot}} \approx \frac{M}{R^2\sqrt{T_{\rm eff}/5777 {\rm K}}} 
\end{equation}
where $\langle \Delta \nu \rangle_{\odot} = 134.9$ \mhz\ and 
$\nu_{\rm max, \odot} = 3,050$ \mhz\ \citep{kje95}, and $R$ and $M$ are in solar
units.
Although the relations are approximate, they have been found to work
quite well, e.g. \citet{bed03,ste08}.
\mlsep\ is the characteristic spacing between consecutive radial overtones
of the same mode degree seen in the power (frequency) spectrum of a 
star with sun-like oscillations (e.g. see Fig.~6 from \citealt{but04}), and 
it is proportional to the square root of the mean density of the star. 
Because it is a repetitive pattern (similar to a periodicity), 
 it is relatively easy to determine from even
low signal/noise data (see e.g. \citealt{hub09,mosapp09,rox09,mat10,ver11} 
who discuss different
methods to determine this value).  
The value of \numax\ is the frequency corresponding to the maximum amplitude
of the bell-shaped amplitude/power spectrum, and it is also a quantity that 
can be more easily observed than, for example, individual oscillation modes.

Because the radii and effective temperatures of these stars are well determined,
the predicted seismic quantities depend only on the mass of the star.
If we substitute directly the derived mass ranges into the equations
then we can predict the range of possible values for these quantities
corresponding to the central model, i.e. not taking into account the changes
in $\alpha$, $Y_{\rm i}$ or \zx.
For \gmb\, we find that for masses = [0.62, 0.64, 0.66] \msol\ 
we calculate \numax\ = [4773, 4927, 5081] \mhz\
and \mlsep\ = [196, 199, 203] \mhz, which correspond to typical periods 
of approximately 4 minutes.   If we can detect these values, even
with poor precision we will still be able to select the optimal mass
range and discard certain solutions.  We note that both $M$ and $Y_{\rm i}$ are
very highly correlated, and so fixing $M$ will present interesting constraints
on $Y_{\rm i}$.  Performing such observations from ground-based 
instrumentation 
should yield successful results
(e.g. a 2+ meter telescope equipped with a highly efficient
and stable spectrograph).
For HD\,122563 we find for $M$ = [0.84, 0.86, 0.88] \msol, 
\numax\ = [5.03, 5.15, 5.26]
\mhz\ and \mlsep\ = [1.05, 1.06, 1.07] \mhz.  
The dominant periods are 
approximately 2.5 days, and observations from ground-based instrumentation 
would be difficult.
In order to use asteroseismic data to help constrain the models
for HD~122563, we would require seismic data from 
space-borne instruments, such as with the CoRoT or {\it Kepler}
missions, to 
provide the necessary precision and determine the individual oscillation modes.

\section{Conclusions}
We have determined the $T_{\rm eff}$, $L$, and $R$ of HD 122563 and \gmb\ 
by using $K$ band interferometric measurements (Table~\ref{tab:observ}) and 
3D/1D limb-darkening for the giant/dwarf.
We find angular diameters of $\theta_{\rm 3D} = 0.940 \pm 0.011$ mas and 
$\theta_{\rm 1D} = 0.679 \pm 0.015$ mas for HD\,122563 and \gmb, respectively,
and these convert into \teff\ = 4598 \pmm\ 41 K for HD\,122563 and 
\teff\ = 4818 \pmm\ 54 K for \gmb.
These new precision temperatures increase 
the well-known difficulty of fitting the  error boxes of
these two metal-poor stars with evolutionary tracks. 
Using the CESAM2k stellar structure and evolution code we found that we 
could match models to the data by using values of the mixing length 
(the parameter $\alpha$) very different from that of the Sun.
We found values of $\alpha$ = 0.68 and 1.31 for the 0.63 \msol\ dwarf 
star and 
the 0.86 \msol\ giant, respectively.  
The order of these values seems consistent with 
recent model analyses \citep{yildiz06,ker0861cyg}.
We found that different equations of state lead to qualitatively but not
quantitively different model parameters for the dwarf star but not for the giant.
The initial helium content comes out similar to the big-bang value, 
the deduced masses are low and their ages are high, consistent with 
expected values for metal-poor halo stars (see Table~\ref{tab:params}).
The masses are determined with a few percent precision and coupling these
with the radii yields well-constrained values of \logg.
For the giant star we found \logg\ = $1.60 \pm 0.04$ 
somewhat higher than the typical
values (1.1 - 1.5) adopted by spectroscopic analyses according to 
the PASTEL catalogue \citep{soubiran10} and for the dwarf star we obtain
\logg\ = 4.59 \pmm\ 0.02 dex.
\citet{barbuy03} determined the O abundance of HD\,122563 assuming two 
different (both justified) values of \logg, and they concluded that their
resulting [O/Fe] = +0.7 
abundance seemed most consistent when they adopt the 
{\it Hipparcos}\footnote{We note that with the new Hipparcos parallaxes the
deduced \logg\~=~1.6.}
\logg\ = 1.5 and not the value determined from ionization equilibrium of 
Fe, \logg\ = 1.1,  a result due possibly to NLTE effects.
This work supports their O determination.
With both \logg\ and \teff\ now very precisely known, these provide 
very important inputs for any spectroscopic analyses, especially for the
determination of neutron-capture element abundances which can constrain 
models of nucleosynthesis.

Finally, we have also predicted the asteroseismic signatures 
\mlsep\ and \numax\ for the two stars and we showed that 
determinations of these quantities for the dwarf star are possible using 
ground-based observations.
For the giant, however, we would require
very long time series in order to resolve the frequency content of the 
oscillations, and this would only be possible with space-borne instruments. 
The asteroseismic data would provide very important constraints because it
would allow us to determine the mass with better precision (using the radius
from this work), and thus the initial helium abundance.

\begin{acknowledgements}

The CHARA Array is funded by
the National Science Foundation through NSF grant AST-0908253 and by Georgia State University through
the College of Arts and Sciences. 
The PTI archival observations of HD\,122563 were collected through the efforts of the PTI Collaboration (\url{http://pti.jpl.nasa.gov/ptimembers.html}). The PTI was operated until 2008 by the NASA Exoplanet Science Institute/Michelson Science Center, and was constructed with funds from the Jet Propulsion Laboratory, Caltech as provided by the National Aeronautics and Space Administration. This work has made use of services produced by the NASA Exoplanet Science Institute at the California Institute of Technology.
This research received the support of PHASE, the high angular resolution
partnership between ONERA, Observatoire de Paris, CNRS and University Denis Diderot Paris 7.
This research made use of the SIMBAD and VIZIER databases at CDS, Strasbourg (France),
and NASA's Astrophysics Data System Bibliographic Services.
The authors acknowledge the role of the SAM collaboration in stimulating this research through regular workshops.
We acknowledge financial support from the ``Programme National de Physique Stellaire" (PNPS) of CNRS/INSU, France.
TSB acknowledges support provided by NASA through Hubble Fellowship grant \#HST- HF-51252.01
awarded by the Space Telescope Science Institute, which is operated by the Association of Universities
for Research in Astronomy, Inc., for NASA, under contract NAS 5- 26555. 
UH acknowledges support from the Swedish National Space Board.
AC is supported in part by an {\it Action de recherche concert\'ee} (ARC) grant from the {\it Direction g\'en\'erale de l'Enseignement non obligatoire et de la Recherche scientifique - Direction de la Recherche scietntifique - Communaut\'e fran\c{c}aise de Belgique.}  AC is also supported by the F.R.S.-FNRS FRFC grant 2.4513.11.
OLC is a Henri Poincar\'e Fellow at the Observatoire de la C\^ote d'Azur.
The Henri Poincar\'e Fellowship is funded by the Conseil G\'en\'eral des Alpes-Maritimes and
the Observatoire de la C\^ote d'Azur.

\end{acknowledgements}

\bibliographystyle{aa}
\bibliography{metal}
\end{document}